\documentstyle[aps, amsmath, amssymb, amsfonts, epsfig, prd, floats,
xspace]{revtex}

\newcommand{\etal}{{\it et al. \@\xspace}}
\newcommand{\eg}{{\it e.g. \@\xspace}}
\newcommand{\ie}{{\it i.e. \@\xspace}}

\makeatletter
\renewcommand\@biblabel[1]{#1.}
\renewcommand{\section}{\@mainheadtrue
\@startsection {section}{1}{\z@}{0.8cm plus1ex minus
 .2ex}{0.5cm plus1ex minus.2ex}{\reset@font\small\bf\noindent}}
\def\subsection{\@mainheadfalse
\@startsection{subsection}{2}{\z@}{0.8cm plus1ex minus
 .2ex}{0.5cm plus1ex minus.2ex}{\reset@font\small\bf\noindent}}
\makeatother

\begin{document}

\title{A dynamical systems approach to geodesics in Bianchi
  cosmologies}  
\author{Ulf S. Nilsson}
\address{Department of Applied Mathematics\\University of Waterloo,
 Waterloo, Ontario, Canada, N2L 3G1}
\author{Claes Uggla}
\address{Department of Engineering Sciences, Physics and Mathematics\\ 
  University of Karlstad, S-651 88 Karlstad, Sweden}
\author{John Wainwright}
\address{Department of Applied Mathematics\\University of Waterloo,
  Waterloo, Ontario, Canada, N2L 3G1}

\maketitle

\begin{abstract}
  To understand the observational properties of cosmological
  models, in particular, the temperature of the cosmic microwave
  background radiation, it is necessary to study their null geodesics.
  Dynamical systems theory, in conjunction with the orthonormal frame
  approach, has proved to be an invaluable
  tool for analyzing spatially homogeneous cosmologies. It is thus
  natural to 
  use such techniques to study the geodesics of these models.
  We therefore augment the Einstein field equations
  with the geodesic equations, all written in dimensionless
  form, obtaining an extended system of   
  first-order ordinary differential equations that simultaneously
  describes the evolution of the gravitational field and the behavior
  of the associated geodesics. It is shown that the
  extended system is a powerful tool for investigating
  the effect of spacetime anisotropies on the temperature 
  of the cosmic microwave background radiation, and that it can also 
  be used for studying geodesic chaos.
\end{abstract}


\section{Introduction}
\label{sec:intro}

The dynamical systems approach to the field equations of general
relativity has been an invaluable tool for gaining qualitative
information about the solution space of the anisotropic but spatially
homogeneous (SH) Bianchi cosmologies (see Wainwright \&
Ellis\footnote{From now on we will refer to this reference as WE.}  
\cite{book:WainwrightEllis1997}
and references therein). In this approach one uses the orthonormal
frame formalism of Ellis \& MacCallum \cite{art:EllisMacCallum1969} to
write the field equations as an autonomous system of first-order
differential equations, the evolution equations for the gravitational
field. One can then apply techniques from the theory of dynamical
systems to obtain qualitative information about the evolution of
Bianchi cosmologies. The essential step is to introduce dimensionless
variables for 
the gravitational field by normalizing with the rate-of-expansion
scalar, or equivalently, the Hubble scalar. A consequence of this
choice of variables is that the equilibrium points of the evolution
equations correspond to self-similar Bianchi models, leading to the
insight that this special subclass of models plays a fundamental role
in determining the structure of the general solution space. An 
added bonus is that the evolution equations are well suited for
doing numerical simulations of Bianchi cosmologies.

In order to understand the observational properties of the
Bianchi models, however, it is necessary to study the behavior of
their 
null geodesics. In this paper we
augment the evolution equations of the gravitational field with the
geodesic  
equations using the components of the tangent 
vector field as the basic variables, thereby creating an extended
system of equations. This yields a system of coupled first-order
ordinary differential 
equations that describes the evolution of the gravitational
field and the behavior of the associated geodesics. It turns out that
normalizing the geodesic 
variables with the energy leads to bounded variables for null and
timelike geodesics, which is of great advantage. 

It is widely believed that a
highly isotropic cosmic microwave background (CMB) temperature implies
that the universe as a whole 
must be highly isotropic about our position, and thus accurately
described by a Friedmann-Lemaitre (FL) model. Bianchi cosmologies
provide an arena for testing this belief. Since the 1960s, various
investigations of the CMB temperature in SH
universes have used the observed anisotropy in the temperature to
place restrictions on the overall anisotropy of the expansion of the
universe, as 
described by the dimensionless scalar\footnote{Here
  $\sigma^2=\tfrac12\sigma_{ab}\sigma^{ab}$ is the norm of the shear
  tensor and $H$ is the Hubble variable.} $\Sigma=\sigma/(\sqrt{3}H)$
(see, for example, Collins \& Hawking
\cite{art:CollinsHawking1973a}). Some of these investigations have
also determined the temperature patterns on the celestial sphere in
universes of different Bianchi types (see, for example, Barrow \etal
\cite{art:Barrowetal1983}). The studies that have
been performed to date, however, suffer from a number of limitations:  
\begin{enumerate}
\item They are restricted to those Bianchi group types that are
  admitted by the FL models. Indeed, the most detailed analyses, for
  example, Bajtlik \etal \cite{art:Bajtliketal1985}, have considered
  only the simplest Bianchi types, namely, I and V. 
  \item The results are derived using linear perturbations of the FL
    models. Such a simple approach cannot be justified in all
    situations
    (see Collins \& Hawking \cite{art:CollinsHawking1973a}, page 316
    and Doroshkevich \etal \cite{art:Doroshkevichetal1975}, page 558).
    \item The analyses provide no bounds on the intrinsic anisotropy
      in the gravitational field, as described, for example, by a
      dimensionless 
      scalar ${\cal W}$ formed from the Weyl curvature tensor (see 
      Wainwright \etal \cite{art:Wainwrightetal1999}, page 2580, for
      the definition of ${\cal W}$).  
\end{enumerate}
The extended system of equations is a powerful tool for investigating
the anisotropy of the CMB temperature free of the above
limitations. In particular, the method can be applied even if the
model in question is not close to an FL model.


The outline of the paper is as follows: In section \ref{sec:extapp} we
show how to extend the orthonormal frame formalism to include the
geodesic equations in SH Bianchi cosmologies. As examples we consider
diagonal class A models and type V and type VI$_h$ 
models of class B. In section \ref{sec:struct} the 
structure of the extended system of equations is discussed. Section
\ref{sec:esiex} contains 
examples of the dynamics of geodesics in some 
self-similar cosmological models.  As a simple non-self-similar
example we consider the locally rotationally symmetric (LRS) Bianchi
type II and I models. Subsequently the Bianchi type IX case is
discussed and   
the  notion of an extended Kasner map for the Mixmaster
singularity is introduced. Section \ref{sec:temp} is devoted to
discussing   
how the extended equations of this paper can be used to analyze the
anisotropies 
of the CMB temperature. We end with a discussion in section
\ref{sec:disc} and mention further possible applications. In Appendix
A we outline how the individual geodesics can be found if needed.

In the paper, latin indices, $a,b,c,...=0,1,2,4$ denote spacetime
indices while greek indices, $\alpha, \beta, ...= 1,2,3$ denote
spatial indices in the orthonormal frame.

\section{Extended orthonormal frame approach}
\label{sec:extapp}



In this section we derive the extended system of first-order
differential equations that governs the evolution of SH universes and
their geodesics. We introduce a group-invariant frame $\left\{{\bf
    e}_a\right\}$ such 
that  ${\bf e}_{0}  
= {\bf n}$ is the unit normal vector field of the SH hypersurfaces.  
The spatial frame vector fields ${\bf e}_{\alpha}$ are then
tangent to these hypersurfaces. The gravitational variables are the
commutation functions of the orthonormal frame, which are customarily
labeled 
\begin{equation}
  \label{eq:basicbasicvar}
\left\{ H, \sigma_{\alpha\beta}, \Omega_\alpha, n_{\alpha\beta},
  a_\alpha\right\}\ ,
\end{equation}
(see WE, equation
(1.63)). The Hubble scalar $H$ describes the
overall expansion of the 
model, $\sigma_{\alpha\beta}$ is the shear tensor and describes the
anisotropy of the expansion, $n_{\alpha\beta}, a_\alpha$ describe the
curvature of the SH hypersurfaces, and $\Omega_\alpha$
describes the 
angular velocity of the frame. The evolution equations for these
variables are given in WE (equations (1.90)-(1.98)). 
To be able to incorporate a variety of sources, we use the standard
decomposition of the 
energy-momentum tensor $T_{ab}$ with respect to the vector field
$\mathbf{n}$, 
\begin{equation}
  T_{ab} = \mu n_an_b + 2q_{(a}n_{b)} + p\left(g_{ab} + n_a n_b\right)
  + \pi_{ab}\ ,
\end{equation}
where
\begin{equation}
  q_an^a = 0\ , \quad \pi_{ab}n^b = \pi_a\!^a = 0 \ , \quad
  \pi_{[ab]} = 0\ .
\end{equation}
Hence, relative to the group invariant frame, we also have the
following source variables
\begin{equation}
  \left\{ \mu, p, q_\alpha, \pi_{\alpha\beta} \right\}\ .
\end{equation}
We now normalize\footnote{See WE, page 112, for the motivation for 
  this normalization.} the gravitational field variables and the
matter variables with the Hubble scalar $H$. We write:  
\begin{equation}
  \label{eq:basicvar}
  \left\{ \Sigma_{\alpha\beta}, R_\alpha, N_{\alpha\beta}, A_\alpha
   \right\} = \left\{
  \sigma_{\alpha\beta}, \Omega_\alpha, 
  n_{\alpha\beta}, a_\alpha, \right\}/H\ ,
\end{equation}
and
\begin{equation}
  \label{eq:mattervar}
  \left\{\Omega, P, Q_\alpha, \Pi_{\alpha\beta}\right\} =
   \left\{\mu, p, q_\alpha, \pi_{\alpha\beta}\right\}/3H^2\ .  
\end{equation}
These new variables are dimensionless and are referred to as
expansion-normalized variables. By introducing a new dimensionless
time variable $\tau$ according to   
\begin{equation}
  \label{eq:taudef}
  \frac{dt}{d\tau}=H^{-1}\ , 
\end{equation}
the equation for $H$ decouples, and can be written as
\begin{equation}
  \label{eq:decq}
  H'= -(1+q)H\ , \quad q = 2\Sigma^2 +
    \tfrac12(\Omega + 3P)\ ,
\end{equation}
where a prime denotes differentiation with respect to $\tau$. The
scalar $\Sigma$ is the dimensionless shear scalar, defined by
\begin{equation}
  \Sigma^2 = \tfrac{1}{6}\Sigma_{\alpha\beta}\Sigma^{\alpha\beta}\ ,
\end{equation}
and $q$ is the deceleration parameter of the
normal congruence of the SH hypersurfaces\footnote{The
  equation for $q$ generalizes equation (5.20) in WE.}. The evolution
equations for 
the dimensionless gravitational field variables follow from equations
(1.90)-(1.98) in WE, using (\ref{eq:basicvar})-(\ref{eq:decq}).

We will now consider the geodesic equations,
\begin{equation}
  \label{eq:geoeq}
  k^a\nabla_a k^b = 0 \ ,
\end{equation}
where $k^a$ is the tangent vector field of the geodesics\footnote{For
  many purposes in SH cosmology, it is sufficient to consider only
  the geodesic tangent vectors, and not the coordinate representation
  of the geodesics themselves. If specific coordinates
  are introduced, the geodesics can be found by the methods outlined in
  appendix A.}. We can regard an individual geodesic as
a curve in a {\em spatially 
homogeneous congruence} of geodesics, in which case the orthonormal
frame components of its tangent vector field satisfy 
\begin{equation}
  \label{eq:SHcongruence}
  {\bf e}_\alpha\left(k^a\right) = 0\ .
\end{equation}
We now use equations (1.15) and (1.59)-(1.62) in WE to write
(\ref{eq:geoeq}) and (\ref{eq:SHcongruence}) in the orthonormal frame
formalism, obtaining
\begin{mathletters}
  \begin{eqnarray}
    k^0\dot{k}^0 &=& -\sigma_{\alpha\beta}k^\alpha k^\beta -
    H\left(k^0\right)^2 \ , \label{eq:k0eq} \\
    k^0\dot{k}^\alpha &=& -k^\beta\left(\sigma^\alpha\!_\beta +
    H\delta^\alpha\!_\beta\right) k^0 +
    \epsilon^\alpha\!_{\beta\nu}n_\mu\!^\nu k^\beta k^\mu - a_\beta
    k^\beta k^\alpha + a^\alpha \left( k_\alpha k^\alpha \right)\ ,
  \end{eqnarray}
\end{mathletters}
where an overdot denotes differentiation with respect to $t$, the
cosmological clock time (synchronous time). We now introduce {\em
  energy-normalized} geodesic variables 
\begin{equation}
  \label{eq:enorm}
  K^\alpha = \frac{k^\alpha}{{\cal E}}\ ,
\end{equation}
where ${\cal E}=k^0$ is the particle energy. The vector $K^\alpha$
satisfies $K_\alpha K^\alpha=1$ for null geodesics, $<1$ for timelike
geodesics, and $>1$ for spacelike geodesics. For null geodesics, the
variables $K_\alpha$ correspond to the {\it direction cosines} of the
geodesic. The equation for the energy ${\cal E}$, equation
(\ref{eq:k0eq}) decouples and can be written as 
\begin{equation}
  \label{eq:decs}
  {\cal E}' = -\left(1+s\right){\cal E} \ ,
\end{equation}
where
\begin{equation}
  s = -1 + K_\alpha
    K^\alpha + \Sigma_{\alpha\beta}K^{\alpha}K^\beta\ . 
\end{equation}
We now summarize the extended system of equations in dimensionless 
form. 
\begin{center}
  {\bf Evolution equations}
\end{center}
\begin{mathletters}
  \label{eq:extsyst}
\begin{eqnarray}
  \Sigma_{\alpha\beta}' &=& -(2-q)\Sigma_{\alpha\beta} + 2\epsilon^{\mu\nu}
\!_{(\alpha}\Sigma_{\beta)\mu}R_\nu - ^3\mathcal{S}_{\alpha\beta}
+\Pi_{\alpha\beta}\ , \label{eq:Sigeq}\\
  N_{\alpha\beta}' &=& qN_{\alpha\beta} + 2\Sigma_{(\alpha}\!^\mu
  N_{\beta)\mu} +
  2\epsilon^{\mu\nu}\!_{(\alpha}N_{\beta)\mu}R_\nu\ , \label{eq:Nabeq} 
  \\ 
  A_\alpha ' &=& qA_\alpha - \Sigma_\alpha\!^\beta A_\beta +
  \epsilon_\alpha \!^{\mu\nu}A_\mu R_\nu\ , \label{eq:Aaeq}\\
  K_\alpha' &=& \left(s + A_\beta K^\beta\right)K_\alpha
  -\Sigma_{\alpha\beta}K^\beta  - \epsilon_{\alpha\beta\gamma}R^\beta
  K^\gamma -
  \epsilon_{\alpha\beta\gamma}N^\gamma\!_\delta K^\delta K^\beta -
A_\alpha\left(K_\beta K^\beta\right)
   \ , \label{eq:Kgeneq}
 \end{eqnarray}
\begin{center}
  {\bf Constraint equations}
\end{center}
\begin{eqnarray}
 \Omega &=& 1-\Sigma^2 - K\ , \\
3Q_\alpha &=& 3\Sigma_\alpha\!^\mu A_\mu -
\epsilon_\alpha\!^{\mu\nu}\Sigma_ \mu\!^\beta N_{\beta\nu}\ ,
\label{eq:Qalphaeq}\\
0 &=& N_\alpha\!^\beta A_\beta \ ,
\end{eqnarray}
\end{mathletters}
where the spatial curvature is given by
\begin{eqnarray}
  ^3\mathcal{S}_{\alpha\beta} &=& B_{\alpha\beta} - \tfrac13\left(
  B_\mu\!^\mu \right)\delta_{\alpha\beta} -
  2\epsilon^{\mu\nu}\!_{(\alpha}N_{\beta)\mu}A_\nu\ , \\
  K &=& -\tfrac{1}{12}B_\mu\!^\mu -A_\mu A^\mu\ ,
\end{eqnarray}
with
\begin{equation}
  B_{\alpha\beta} = 2N_\alpha\!^\mu N_{\mu\beta} - \left( N_\mu\!^\mu \right)
N_{\alpha\beta}\ .
\end{equation}
Accompanying the above system of equations are, if necessary, 
equations for matter variables. For example, if the
source were a tilted perfect fluid, additional equations for the
tilted fluid 4-velocity would have to be added.

Note that the null geodesics, characterized by $K_\alpha K^\alpha=1$,
define an invariant 
subset. This is easily seen from the auxiliary equation for the length
of the vector $K^\alpha$,
\begin{equation}
  \label{eq:klength}
  \left(K_\alpha K^\alpha\right)' = -2\left(1-K_\alpha K^\alpha\right)\left(
1+s\right)\ .
\end{equation}
From now on we will restrict our considerations to null geodesics, in
which case the expression for $s$ simplifies to
\begin{equation}
  \label{eq:sgeneral}
  s=\Sigma_{\alpha\beta}K^\alpha K^\beta\ .
\end{equation}

\subsection{Examples: Some non-tilted perfect-fluid models}
\label{sec:diag}

For non-tilted perfect fluid models, the 4-velocity of the fluid,
${\bf u}$, coincides with the normal vector field ${\bf n}$ and
$Q_a=\Pi_{ab}=0$. It will also be assumed that the cosmological fluid  
satisfies a linear barotropic scale-invariant equation of state,
$p=(\gamma-1)\mu$, or equivalently, $P=(\gamma-1)\Omega$, where
$\gamma$ is a 
constant. From a physical point of view, the most important values are  
$\gamma=1$ (dust) and $\gamma=\tfrac{4}{3}$ (radiation). The value
$\gamma=0$ corresponds to a cosmological constant and the value
$\gamma=2$ to a ``stiff fluid''. Here it is assumed that
$0\leq\gamma\leq2$. Our focus will be on diagonal Bianchi
models. These are the class A models, and the $N_\alpha\!^\alpha=0$
models of class B, \ie models of type V and special models of type
VI$_h$ (see Ellis \& MacCallum \cite{art:EllisMacCallum1969}).

\subsubsection*{Class A models}
\label{sec:classA}
For the class A models, $A_\alpha=0$, it is possible to choose a
frame such that $N_{\alpha\beta} = {\rm diag}(N_1,N_2,N_3)$,
$R_\alpha = 0$, and   
\begin{equation}
  \Sigma_{\alpha\beta} =
  \mathrm{diag}(\Sigma_++\sqrt{3}\Sigma_-\ , \Sigma_+-\sqrt{3}
  \Sigma_-\ , -2\Sigma_+)\ ,
\end{equation}
(see WE, page 123). Here we have chosen to adapt the
decomposition of the trace-free 
shear tensor $\Sigma_{\alpha\beta}$ to the third direction, rather
than the first direction, as in WE. The anisotropic spatial curvature
tensor $^3{\cal S}_{\alpha\beta}$ is also diagonal and we label its
components in an analogous way:
\begin{equation}
  ^3{\cal S}_{\alpha\beta} = {\rm diag}\left( {\cal S}_+ +
    \sqrt{3}{\cal S}_- \ , {\cal S}_+ - \sqrt{3}{\cal S}_- \ ,
    -2{\cal S}_+ \right) \ .
\end{equation}
With the above choice of frame, (\ref{eq:extsyst}) leads
to an extended system of equations of the form:
\begin{center}
  {\bf Evolution equations}
\end{center}
\begin{mathletters}
  \label{eq:classAeq}
\begin{eqnarray}
  \Sigma_\pm' &=& -(2-q)\Sigma_\pm - \mathcal{S}_\pm \ , \\
  N_1' &=& (q+2\Sigma_++2\sqrt{3}\Sigma_-)N_1\ ,\\ 
  N_2' &=& (q+2\Sigma_+-2\sqrt{3}\Sigma_-)N_2\ ,\\
  N_3' &=& (q-4\Sigma_+)N_3\ , \\
  K_1' &=& (s-\Sigma_+ - \sqrt{3}\Sigma_-)K_1 + (N_2-N_3)K_2K_3\ ,
  \label{eq:K1A}\\
  K_2' &=& (s-\Sigma_+ + \sqrt{3}\Sigma_-)K_2 + (N_3-N_1)K_1K_3\ ,
  \label{eq:K2A}\\ 
  K_3' &=& (s+2\Sigma_+)K_3 + (N_1-N_2)K_1K_2 \label{eq:K3A}\ ,
\end{eqnarray}
\end{mathletters}
where 
\begin{align}
  q &= \tfrac12(3\gamma-2)(1-K) +
  \tfrac{3}{2}(2-\gamma)(\Sigma_+^2+\Sigma_-^2)\ , \\
  s &= (1-3K_3^2)\Sigma_+ + \sqrt{3}\left(K_1^2 - K_2^2\right)\Sigma_-
  \ , \label{eq:sclassA}\\ 
  \mathcal{S}_+ &= \tfrac{1}{6}\left[ (N_1-N_2)^2 - N_3(2N_3-N_1-N_2)
  \right]\ ,\\
  \mathcal{S}_- &= \tfrac{1}{2\sqrt{3}}(N_2-N_1)(N_3-N_1-N_2) \ , \\
  K &= \tfrac{1}{12}\left[ N_1^2 + N_2^2 + N_3^2 - 
    2\left(N_1N_2 + N_2N_3 + N_3N_1\right)\right]\ .
\end{align}
The density parameter $\Omega$ is defined by
\begin{equation}
  \Omega = 1-\Sigma_+^2 - \Sigma_-^2 - K\ .
\end{equation}

\subsubsection*{Diagonal class B models}
\label{sec:diagB}

For the non-exceptional class B models with $n_\alpha\!^\alpha=0$
(denoted Ba and Bbi in Ellis \& MacCallum
\cite{art:EllisMacCallum1969}, pages 115,121-122), we can choose the
spatial frame vectors ${\bf e}_\alpha$ so that the shear tensor
$\Sigma_{\alpha\beta}$ is diagonal, $R_\alpha=0$,
$A_\alpha=(0,0,A_3)$, and the only 
non-zero components of $N_{\alpha\beta}$ are $N_{12}=N_{21}$.
These models correspond to Bianchi type V and
special type VI$_h$ models. Equations (\ref{eq:Nabeq}) and
(\ref{eq:Aaeq}) imply that $(N_{12}/A_3)'=0$, \ie we can write
\begin{equation}
  \label{eq:hdef}
  A_3^2=-hN_{12}^2\ ,
\end{equation}
where $h$ is the usual class B group parameter.
For convenience, we introduce a new parameter $k$ according to 
\begin{equation}
k = \frac{1}{\sqrt{1-3h}}\ .
\end{equation}
The type V models are characterized by $k=0$, while $k=1$ corresponds
to type VI$_0$ models, which are actually of Bianchi class A.
Equation (\ref{eq:Qalphaeq}) leads to restrictions on the shear tensor 
$\Sigma_{\alpha\beta}$, which can be written as
\begin{equation}
  \label{eq:diagsigma}
  \Sigma_{\alpha\beta} =
  {\rm diag}\left(-k+\sqrt{3}\sqrt{1-k^2}\ ,\quad
  -k-\sqrt{3}\sqrt{1-k^2}\ , \quad  
  2k\right)\Sigma_\times\ .
\end{equation}
We now introduce a new variable $A$, and rewrite (\ref{eq:hdef}) in
terms of $k$, obtaining
\begin{equation}
  \label{eq:newAdef}
  A_3=\sqrt{1-k^2}A\ , \quad N_{12}=\sqrt{3}kA \ .
\end{equation}
Using (\ref{eq:diagsigma}) and (\ref{eq:newAdef}), the extended system
(\ref{eq:extsyst}) reduces to the following set:
\begin{center}
  {\bf Evolution equations}
\end{center}
\begin{mathletters}
  \label{eq:typeVIhequations}
\begin{eqnarray}
  \Sigma_\times' &=&  -(2-q)\Sigma_\times - 2kA^2 \ , \\
  A' &=& (q+2k\Sigma_\times)A\ , \\
  K_1' &=& \left[s - (k-\sqrt{3}\sqrt{1-k^2})
    \Sigma_\times + \left(\sqrt{1-k^2} +
      \sqrt{3}k\right)AK_3\right]K_1\ , \\ 
  K_2' &=& \left[s - (k+\sqrt{3}\sqrt{1-k^2})
    \Sigma_\times + \left(\sqrt{1-k^2} -
      \sqrt{3}k\right)AK_3\right]K_2\ , \\ 
  K_3' &=& \left[s+\sqrt{1-k^2}AK_3 + 2k\Sigma_\times\right]K_3 
  -\left[\sqrt{3}(K_1^2 - K_2^2)k +
    \sqrt{1-k^2}\right]A\ , 
\end{eqnarray}
\end{mathletters}
where
\begin{align}
 q &= \tfrac{3}{2}(2-\gamma)\Sigma_\times^2 + 
\tfrac12(3\gamma-2)(1-A^2)\ , \\
 s &= \left[ (k-\sqrt{3}\sqrt{1-k^2})K_1^2 +
   (k+\sqrt{3}\sqrt{1-k^2})K_2^2  
        +2kK_3^2 \right]\Sigma_\times \label{eq:sdefA}\ .
\end{align}
The density parameter $\Omega$ is given by
\begin{equation}
\Omega = 1-\Sigma_\times^2 - A^2\ .
\end{equation}

\section{Structure of the extended system of equations}
\label{sec:struct}

We now give an overview of the structure of the combined system of
gravitational and geodesic equations. For simplicity, we only consider
the non-tilted perfect fluid models described in section
\ref{sec:extapp}. The basic 
dimensionless variables are
\begin{eqnarray}
  {\bf x} &=& \left\{ \Sigma_{\alpha\beta}\ , \  N_{\alpha\beta}\ ,
    \ A_\alpha\right\}\ , \\
  {\bf K} &=& \left\{K_\alpha\right\}\ .
\end{eqnarray}
We have shown that the Einstein field equations lead to an autonomous
system of differential equations of the form
\begin{equation}
  \label{eq:feq}
  {\bf x}' = {\bf f}\left({\bf x}\right)\ ,
\end{equation}
(see (\ref{eq:classAeq}a-d) and (\ref{eq:typeVIhequations}a-b)). 
The geodesic equations lead to an autonomous system of 
differential equations of the form 
\begin{equation}
  \label{eq:geeq}
  {\bf K}' = {\bf h}\left( {\bf x}, {\bf K} \right)\ ,
\end{equation}
which is coupled to (\ref{eq:feq}) (see (\ref{eq:classAeq}e-g) and
(\ref{eq:typeVIhequations}c-e)). The geodesic variables 
$K_\alpha$ also satisfy the constraint
\begin{equation}
  \label{eq:nullcons}
  K_\alpha K^\alpha=1\ ,
\end{equation}
and hence define a 2-sphere, which we 
will call the {\it null sphere}. In the context of cosmological
observations, one can identify the null sphere with the celestial
2-sphere. We will refer to the entire 
set, equations (\ref{eq:feq})-(\ref{eq:nullcons}) 
for ${\bf x}$ and ${\bf K}$, as the extended  
scale-invariant system of evolution equations, or
briefly, {\it the extended system of equations.}

There are also two variables with dimension, namely the Hubble scalar
$H$ and the particle energy ${\cal E}$. These scalars satisfy the decoupled
equations (\ref{eq:decq}) and (\ref{eq:decs}). They are thus 
determined by quadrature once a solution of the extended system of
equations has been found.

We now discuss the structure of the state
space of the extended system of equations
(\ref{eq:feq})-(\ref{eq:geeq}). The fact that the gravitational field
equations (\ref{eq:feq}) are independent of ${\bf K}$ implies that the
state space has a product structure, as follows. For models of a
particular 
Bianchi type the gravitational variables ${\bf x}$ belong to a
subset ${\cal B}$ of ${\Bbb R}^n$ (see WE, section 
6.1.2 for Bianchi models of class A). Because of the constraint 
(\ref{eq:nullcons}), the
extended state space is the Cartesian product ${\cal B}\times
S^2$, where $S^2$ is the 2-sphere. The 
orbits in ${\cal B}$ lead to a decomposition of the extended state
space into 
a family of invariant sets of the form $\left\{\Gamma\right\}\times
S^2$, where $\Gamma$ is an orbit in ${\cal B}$. Given a cosmological
model $U$, its evolution is described by an orbit $\Gamma_U$ in ${\cal
  B}$. The orbits in the invariant set $\left\{\Gamma_U\right\}\times
S^2$ then describe the evolution of the model and all of its null
geodesics. We shall refer to $\left\{\Gamma_U\right\}\times S^2$ as
the {\it geodesic submanifold} of the model $U$ in the extended state
space ${\cal B}\times S^2$. In physical terms, with the null sphere
representing the celestial 
sky, the geodesic submanifold of a model $U$ determines the anisotropy
pattern of the CMB temperature in the model $U$ (see section
\ref{sec:temp}).  

An advantage of using a scale-invariant formulation of the
gravitational evolution equations is that models admitting an
additional homothetic 
vector field, the so-called {\em self-similar models}, appear as
equilibrium points (see WE, page 119). The equilibrium points of the
field equations are  
constant vectors ${\bf x}={\bf x}_0$ satisfying
${\bf f}\left({\bf x}_0\right)=0$, where ${\bf f}$ is the
function in (\ref{eq:feq}). In this case, the geodesic
equations,
\begin{equation}
  \label{eq:selfsimgeo}
{\bf K}' = 
{\bf h}({\bf x}_0, {\bf K})\ ,
\end{equation} 
form an independent autonomous system of differential
equations. The equilibrium points of the extended system
(\ref{eq:feq})-(\ref{eq:geeq}) are points
$\left({\bf x}_0,{\bf K}_0\right)$ that satisfy
\begin{equation}
  {\bf f}({\bf x}_0) = {\bf 0} \ ,  
  \quad {\bf h}({\bf x}_0, 
  {\bf K}_0) = {\bf 0}\ .
\end{equation}
Knowing the equilibrium points of the field equations (see WE, section
6.2 for the class A models) one simply has to 
find the equilibrium points of the geodesic equations in 
(\ref{eq:selfsimgeo}). The fixed point theorem for the sphere
guarantees  
that the system of geodesic equations for self-similar models has at
least one equilibrium point on the 
null sphere. Since the null sphere can be identified with the celestial
2-sphere, equilibrium points of the extended system of equations
correspond to the existence of geodesics in fixed directions,
\ie purely ``radial'' geodesics.

\section{Examples of extended dynamics}
\label{sec:esiex}

In this section we will consider some examples of self-similar and
non-self-similar models. For self-similar models, the extended
system of equations reduces to (\ref{eq:selfsimgeo}),
and it is possible to visualize the dynamics of the geodesics. The
most important self-similar models are those of Bianchi type I and II,
namely the flat Friedmann-Lemaitre model, the Kasner 
models and the Collins-Stewart LRS type II 
model (see Collins \& Stewart \cite{art:CollinsStewart1971}), since
these models 
influence the evolution of models of more 
general Bianchi types. For non-self-similar 
models, the dimension of the extended system of equations is usually too
large to permit a complete visualization of the dynamics although one
can apply the standard techniques from the theory of dynamical
systems. In the simplest SH cases, however, one can visualize the
dynamics, and as an example of non-self-similar extended dynamics, we
will consider the Bianchi type II LRS models. We will end the section
with a discussion of the Bianchi type IX models.

\subsection{Self-similar models}

\subsubsection*{The flat Friedmann-Lemaitre model}
The flat FL model corresponds to the following
invariant subset of the extended system of equations for class A
models: 
$\Sigma_+ = \Sigma_- = N_1 = N_2 = N_3 = 0$. The remaining
equations in (\ref{eq:classAeq}) are just 
\begin{equation}
K_\alpha' = 0 \ , \quad \alpha=1,2,3\ .
\end{equation}
Thus, all orbits corresponding to null geodesics are equilibrium
points and the null sphere is an equilibrium set. This fact implies
that all null geodesics are radial geodesics. 

\subsubsection*{Kasner models}
Although these are vacuum models, they are extremely
important since they are asymptotic states for many of the more general
non-vacuum models. The models correspond to the 
Bianchi type I invariant vacuum subset of the extended
system of  equations for the class A models: $N_1=N_2=N_3=0 \ ,
\Sigma_+^2 + \Sigma_-^2 = 1$, where $\Sigma_+$ and $\Sigma_-$ are 
constants. 

The remaining equations are the geodesic equations (\ref{eq:K1A}) --
(\ref{eq:K2A}) with $N_1=N_2=N_3=0$ and with $s$ given by  
(\ref{eq:sclassA}). We note that these equations are invariant under the
discrete transformations
\begin{equation}
  \left( K_1, K_2, K_3 \right) \rightarrow \left( \pm K_1, \pm K_2,
  \pm K_3 \right)\ .
\end{equation}
The constant values of $\Sigma_+$ and $\Sigma_-$ determine the
so-called {\it Kasner parameters} $p_\alpha$ according to (see
WE, equation (6.16) with 1,2,3 relabeled as 3,1,2)
\begin{equation}
  p_{1,2} = \tfrac13\left(1+\Sigma_+\pm\sqrt{3}\Sigma_-\right)\ ,
  \quad p_3 = \tfrac13\left(1-2\Sigma_+\right)\ .
\end{equation}
One can also
label the Kasner solutions using an angle $\varphi$, defined by
$\Sigma_+=\cos\varphi$ and $\Sigma_- = \sin\varphi$.
All distinct models are obtained when $\varphi$ assumes the values
$0\leq\varphi\leq\tfrac{\pi}{3}$.
The equilibrium points for these equations are listed in table 1,
together with their eigenvalues. In the LRS cases 
$p_1=p_2\neq 0$ there is a circle $C_{12}$ of equilibrium points.

\begin{table}
  \label{tab:kasner}
  \begin{tabular}{@{}ccll}
      \mbox{} & Eq. point & $K_1,K_2,K_3$ & Eigenvalues\\
      \tableline
       & $P_1^\pm$ & $(\pm 1,0,0)$ & $-3(p_2-p_1)\ , \ -3(p_3-p_1)$ \\ 
        \cline{2-4}  
         $p_1\neq p_2 \neq p_3$ & $P_2^\pm$ & $(0,\pm 1,0)$ &
      $-3(p_3-p_2)\ , 
      \ -3(p_1-p_2)$ \\ 
      \cline{2-4} 
         & $P_3\pm$ &
        $(0,0,\pm 1)$ & $-3(p_2-p_3)\ , \ -3(p_1-p_3)$ \\
        \cline{2-4}
        \tableline
         & $C_{12}$ & $(\cos\psi, \sin\psi, 0)$
        & $ 0\ , \  3$ \\
        \cline{2-4}
        \raisebox{2.0ex}[0cm][0cm]{$p_1=p_2 \neq 0$}& $P_3^\pm$ & 
        $(0,0,\pm 1)$  
        & $-3 \ , \ -3$ \\
        \cline{2-4}
        \tableline
         & $C_{12}$ & $(\cos\psi, \sin\psi,0)$  
        & $ 0\ , \  -3$  \\
        \cline{2-4}
        \raisebox{2.0ex}[0cm][0cm]{$p_1=p_2=0$} & $P_3^\pm$ & 
      $(0,0,\pm 1)$    
        & $3 \ , \ 3$ \\ 
      \end{tabular}
      \caption{The equilibrium points of the geodesic equations for Kasner 
        models, written in terms of $p_1$, $p_2$, and $p_3$ as defined in
        the text. The parameter $\psi$ is a constant satisfying
        $0\leq\psi\leq2\pi$. The
        eigenvalues for other LRS models than $p_1=p_2$ can be found by
        appropriate permutations.}
\end{table}

For each set of Kasner parameters $p_1,p_2$, and $p_3$ the geodesic
equations admit local
sinks and local sources, which can be identified by considering the
signs of the eigenvalues in table 1. It turns out that
these local sinks/sources are in fact global, \ie attract/repel all
orbits, and hence define the future/past attractor. The reason for
this is the existence of monotone functions that force {\it 
  all} orbits to approach the local sinks/sources into the
future/past. For example, for models with $\Sigma_+\neq0$ we have the
function
\begin{equation}
  Z = \frac{K_3^2}{K_1K_2}\ , \quad Z' = 3(1-3p_3)Z\ .
\end{equation}
The future and past attractors are listed in table 2 for the three  
cases $p_1=p_2\neq0$ (\ie $\varphi=0$), $p_1>p_2>0>p_3$
(\ie $0<\varphi<\tfrac{\pi}{3}$) and $p_2=p_3=0$
(\ie $\varphi=\tfrac{\pi}{3}$).

\begin{table}
  \label{tab:kasner2}
  \begin{tabular}{@{}ccc}
      Kasner parameters & Past attractor & Future attractor \\
      \tableline
      $p_1>p_2>0>p_3$ & $\left\{ P_1^+ \cup P_1^-\right\}$ & $\left\{
        P_3^+ \cup P_3^-\right\}$ \\
      \tableline
      $p_1=p_2\neq0$ & $C_{12}$ & $\left\{
        P_3^+ \cup P_3^-\right\}$ \\
      \tableline
      $p_2=p_3=0$ & $\left\{ P_1^+ \cup P_1^- \right\}$ & $C_{23}$ \\
    \end{tabular}
  \caption{The past and future attractors in the state space for null
      geodesics in the Kasner models whose parameters 
    satisfy $p_1 \geq p_2 \geq  0 \geq p_3$. The results for other
    ordering of parameters can be obtained by appropriate permutations.}
\end{table}

\begin{figure}[ht]
  \centerline{ \hbox{
      \epsfig{figure=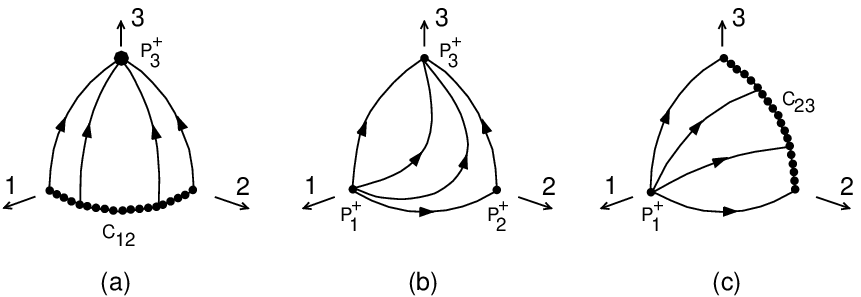, width=0.6\textwidth}}}
  \caption{The dynamics of null geodesics in the Kasner models, in the
      cases (a) $p_1=p_2=\tfrac{2}{3}, p_3=-\tfrac{1}{3}$
      ($\varphi=0$), (b) $p_1 > p_2 > 0 > p_3$
      ($0<\varphi<\tfrac{\pi}{3}$) and (c) $p_1=1, p_2=p_3=0$
      ($\varphi=\tfrac{\pi}{3}$).}
  \label{fig:kasner1}
\end{figure}

In figure \ref{fig:kasner1} we show the orbits
corresponding to null geodesics in the Kasner models for the three
cases $p_1 = p_2 \neq 0$ ($\varphi=0$), $p_1>p_2>0>p_3$, and
$p_2=p_3=0$ ($\varphi=\tfrac{\pi}{3}$). Due to symmetry, it is
sufficient to show the subset of the null sphere defined by $K_1, K_2,
K_3\geq 0$.

\subsubsection*{The Collins-Stewart LRS type II solution}
The Collins-Stewart model corresponds to  the following
  submanifold\footnote{Note the
  incorrect numerical factor on page 
131 in WE.} of the extended system
  of equations:
\begin{equation}
  \Sigma_+ = \tfrac{1}{8}(3\gamma-2)\ , \ \Sigma_- = 0 \ , \ N_1 = N_2
  = 0 \ , \ N_3 = \tfrac{3}{4}\sqrt{(2-\gamma)(3\gamma-2)}\ ,
\end{equation}
with $\tfrac{2}{3} < \gamma < 2$. Due to the symmetries, we need only 
consider $K_3\geq0$. The 
equilibrium points and sets are listed in table 3.
\begin{table}
  \label{tab:typeII}
  \begin{tabular}{@{}cll}
    Eq. point & $K_1, K_2, K_3$ & Eigenvalues \\
    \tableline
    $P_3$ & $0,0,1$ &
  $\tfrac{3}{2}(2-\gamma), 
  -\tfrac{3}{8}\left( 3\gamma-2 \pm 2ib \right)$ \\
  \tableline
  $C_{12}$ & $\cos\psi, \sin\psi,0$ &  
  $\tfrac{3}{4}(2+\gamma),\tfrac{3}{8}(3\gamma-2),0$ \\
\end{tabular}
  \caption{The equilibrium points and sets for the null geodesic
    equations in the Collins-Stewart LRS type II solution. The
  parameter $\psi$ 
  satisfies $0\leq\psi \leq 2\pi$. Note that two of
  the eigenvalues for the equilibrium point ${\cal C}_{12}$ are
  complex. The constant $b$ is 
  given by $b = \sqrt{ (3\gamma-2)(2-\gamma) }$.}
\end{table}

\begin{figure}[ht]
  \centerline{ \hbox{
      \epsfig{figure=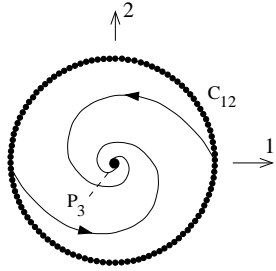, width=0.2\textwidth}}}
  \caption{The dynamics of the null sphere for the Collins-Stewart
      LRS type II model, as viewed from the positive 3-axis.}
  \label{fig:IIgeo}
\end{figure}

The equilibrium set $C_{12}$ is the source, while the
the equilibrium point $P_3$ is a stable focus. Note that $K_3$ is an
increasing monotone function. The 
dynamics of the null geodesics is shown in figure
\ref{fig:IIgeo}. Note that 
there are no changes in the stability of the equilibrium points for
$\tfrac{2}{3}<\gamma<2$.

\subsection{Non--self--similar models}

The previous examples are simple in the sense that we only had to
consider the geodesic part of the extended system of equations. For
non--self--similar models, the full system has to be considered, which
means 
that the dynamics will in general be difficult to visualize due to the
high dimensions of the extended state space. To illustrate the ideas,
we consider the null geodesics in Bianchi type I and II LRS models. The
behavior of geodesics in the Mixmaster model is also discussed.

\subsubsection*{LRS Bianchi type I and II models}
The type II LRS models correspond to the invariant subset
$\Sigma_- = 0$, 
$N_1=N_2=0$ of the extended system of equations (\ref{eq:classAeq}) for the 
class A 
models, while the type I models, in addition, require $N_3=0$.
For null geodesics, the extended system is five dimensional (four for
type I), with one constraint $K_\alpha K^\alpha=1$. Defining 
\begin{equation}
  \label{eq:K1K2}
  K_1 = R\cos\chi\ , \ K_2 = R\sin\chi\ ,
\end{equation}
where
\begin{equation}
  R=\sqrt{1-K_3^2}\ ,
\end{equation}
leads to a decoupling of the $\chi$-equation,
\begin{equation}
  \chi' = N_3K_3\ ,
\end{equation}
leaving a reduced extended system
\begin{mathletters}
  \label{eq:IILRS}
\begin{eqnarray}
  \Sigma_+' &=& -(2-q)\Sigma_+ + \tfrac{1}{3}N_3^2\ , \label{eq:IIsig}\\
  N_3' &=& (q-4\Sigma_+)N_3\ , \label{eq:IIN3}\\
  K_3' &=& 3\Sigma_+(1-K_3^2)K_3\label{eq:IIK3}\ ,
\end{eqnarray}
\end{mathletters}
with
\begin{equation}
  q = \tfrac12(3\gamma-2)(1-\tfrac{1}{12}N_3^2) +
  \tfrac{3}{2}(2-\gamma)\Sigma_+^2\ .
\end{equation}

\begin{figure}[ht]
  \centerline{ \hbox{
      \epsfig{figure=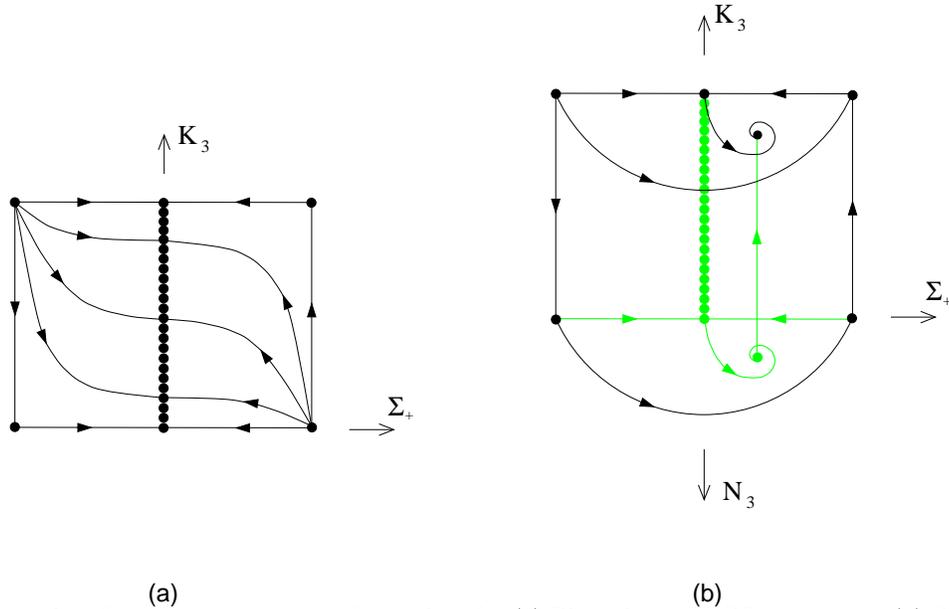, width=0.7\textwidth}}}
  \caption{The dynamics of the extended system of equations for (a)
      Bianchi type I LRS models and (b) Bianchi type II LRS models.}
  \label{fig:IILRS}
\end{figure}

The state space associated with (\ref{eq:IILRS})
is the product set ${\cal
  B}\times\left[0,1\right]$, where ${\cal B}$ is the state space of
the Bianchi type II LRS cosmologies (or type I, in the case $N_3=0$),
associated with the subsystem of 
(\ref{eq:IILRS}a)-(\ref{eq:IILRS}b). In this representation the
null sphere is replaced by the single geodesic variable $K_3$,
with $0\leq K_3\leq1$. The remaing two geodesic variables are given
by (\ref{eq:K1K2}). We refrain from giving the various
equilibrium points and their eigenvalues. Instead we give the
three-dimensional extended state space of (\ref{eq:IILRS}) in
figure \ref{fig:IILRS}b  
and the two-dimensional invariant set $N_3=0$ in figure
\ref{fig:IILRS}a. In figure \ref{fig:IILRS}b we have simply shown the
skeleton of the state space, \ie the equilibrium points and the
various heteroclinic orbits that join the equilibrium points. The
figures depict the situation when $\tfrac{2}{3}<\gamma<2$ 
since there are no bifurcations for this interval. The sources and
sinks can be deduced from the figures.
A detailed picture of the orbits in the gravitational state space
$K_3=0$ is given in WE (see figure 6.5). We 
note that the orbits in the invariant set $K_3=1$ are identical to
those with 
$K_3=0$. Knowing the orbits in $K_3=0$ and $K_3=1$, one can visualize
the structure of the geodesic submanifolds -- they are vertical
surfaces of the form $\left\{\Gamma\right\}\times\left[0,1\right]$,
where $\Gamma$ is an orbit in the subset $K_3=0$.

\subsubsection*{Comments on Bianchi type IX models}
It was recognized a long time ago that the oscillatory approach to the
past or  
future singularity of Bianchi IX vacuum models, the so-called
Mixmaster attractor, displays random features, see 
\eg Belinskii \etal \cite{art:BKL}, and hence is a potential
source of chaos.  
This behavior is also expected in non-vacuum Bianchi models with
various 
matter sources (see section 6.4.1 in WE and references therein).  
Numerical studies of the governing equations of vacuum
Bianchi IX models toward the initial singularity have shown that the
variables 
$\Sigma_\pm$ and the $N_\alpha$ remain bounded. These studies have also 
shown that the
projection of the orbits onto the $\Sigma_\pm$-plane is given, at least to a 
high accuracy, by the Kasner map (see WE, section 11.4.2). The
transition between two 
different Kasner states is 
described by a vacuum Bianchi type II orbit except when the Kasner
state is close to an LRS Kasner model where this approximation is no
longer valid. When discussing the Mixmaster attractor, one is usually
discussing individual orbits. Thus a corresponding discussion for the
extended system implies a discussion about the Bianchi type IX
geodesic submanifold. 
Precisely as an individual Bianchi type IX orbit can be
approximated by a sequence of Bianchi type II orbits, one can
approximate a type 
IX geodesic submanifold with a sequence of type II geodesic
submanifolds. The stable equilibrium points within the type II
geodesic 
submanifold reside in the type I geodesic boundary submanifold of
these models and correspond to geodesics in the 1,2 or 3 directions,
{\em modulo sign}, depending on the particular Kasner point. 

We will only consider such sequences of Kasner states
for which the Kasner models are not close to any LRS models.
As the evolution progresses, the $\tau$-time that the system spends
close to a 
Kasner state, a so-called Kasner epoch, becomes successively longer and
should thus be well described by the appropriate equilibrium point.
If we
assume that during a certain Kasner epoch the qualitative behavior of 
a geodesic is given by the stable equilibrium point of the
extended system of equations for these models, we can extend the
Kasner map to 
include 
the stable direction of the geodesic. Since we are excluding the LRS
Kasner models there will never appear any equilibrium sets as they
only when $\varphi$ is a multiple of
$\pi/3$. The direction of stability, modulo sign, 
is given in table 4 as a 
function of $\varphi$. These results follow from the general stability
of the equilibrium points given in table 1, by changing
the signs of the eigenvalues since the models are approaching the the
initial singularity, \ie $\tau\rightarrow-\infty$. 

\begin{table}
  \label{tab:Geostable}
      \begin{tabular}{@{}lc}
        Range of $\varphi$ & Stable geodesic direction \\
        \tableline
        $0<\varphi<2\pi/3$ & $1$ \\
        \tableline
        $2\pi/3<\varphi<4\pi/3$ & $3$ \\
        \tableline
        $4\pi/3<\varphi<2\pi$ & $2$ \\
      \end{tabular}
      \caption{The stable geodesic direction for the geodesics for a
        particular Kasner  
        epoch for different $\varphi$'s. By assumption we exclude all
        the LRS Kasner models, \ie models when $\varphi$ is a multiple
        of $\pi/3$.}
    \end{table}

\begin{figure}[ht]
  \centerline{ \hbox{
      \epsfig{figure=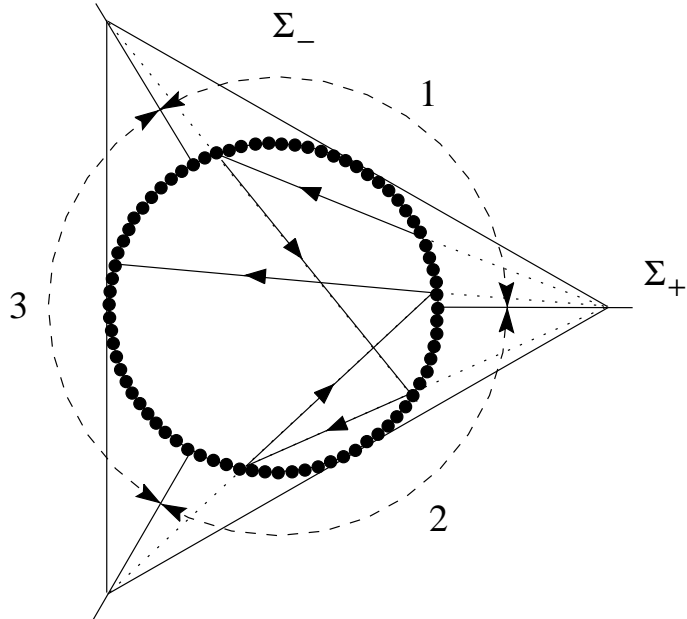}}}
  \caption{The extended Kasner map including the stable geodesic 
        directions. The sequence of Kasner states depicted has the
        following sequence of stable geodesic directions: 1,1,2,2,1,3.}
  \label{fig:extended}
\end{figure}

Starting with a geodesic whose tangent
vector satisfies $K_1, K_2, K_3 \geq 0$ in a given Kasner epoch with 
$0<\varphi_{\text{ini}}<\pi/3$, the stable geodesic direction is the
1--direction. The system
then evolves, according to the Kasner map, into a state with
$\tfrac{\pi}{3}<\varphi_{\text{fin}}<\pi$. 
Depending on the initial value $\varphi_{\text{ini}}$, the stable geodesic
direction can either stay the same
($\text{arccos}\left(\tfrac{13}{14}\right)<\varphi_{ini}<
\tfrac{\pi}{3}$) or change to the 2-direction
($0<\varphi_{\text{ini}}<\text{arccos}\left(\tfrac{13}{14}\right)$).
This process is then repeated as the state changes again. This
extended Kasner map is shown in figure  
\ref{fig:extended}. In the figure, a whole sequence of Kasner states
is also shown where the stable geodesic directions are given by the
sequence  
$1,1,2,2,1,3$. 
The above discussion of the behavior of geodesics toward the Mixmaster
singularity is based on the assumption that as the system 
changes from one Kasner epoch to another, the geodesics are not affected. 
This means that a tangent vector to the geodesic with $K_{1,2,3}>0$
can never  
evolve into a tangent vector with one or more of the $K_\alpha$'s
negative.

This assumption is rather crude since the change of Kasner epochs is
approximately described by 
a vacuum Bianchi type II orbit, for which the geodesic equations, if viewed 
as separate from the field equations, are non-autonomous. Taking this into 
account limits the predictability of the ``extended Kasner map'' in that it 
cannot predict if a geodesic evolves in the positive or negative direction
of the stable geodesic direction. We also note that it is only in
$\tau$-time that the system spends longer and 
longer time in each Kasner epoch. In synchronous time, the interval
becomes shorter and shorter.

From the above discussion, it is
expected that there will be some kind of geodesic chaos in the
development toward the  
initial and final singularity. To substantiate this, one would need,
in addition to further analytical results, careful numerical
studies. We believe that the extended system of equations, as
presented in this paper, may be very well suited for  
such an analysis. The next step would be to study the extended system of
equations for Bianchi type II vacuum models.

\section{Temperature distribution}
\label{sec:temp}

In this section we describe how the extended system of equations
(\ref{eq:feq})-(\ref{eq:nullcons}), together with the decoupled
energy equation (\ref{eq:decs}), can
be used to study the temperature of the CMB in an SH universe. We
regard the photons of the CMB as a test fluid, \ie one which is
not a source of the gravitational field. It is possible
to include the effect of the CMB photons on the gravitational field
by considering two non-interacting fluids, radiation and dust, using
the approach of Coley \& Wainwright \cite{art:ColeyWainwright1992}. We
will not  
do this since the effects of the radiation fluid is not expected to
change our results significantly. To obtain 
the present temperature of the CMB, the photon energies are
integrated along the null geodesics connecting points of emission on
the surface of last scattering to the event of observation at the
present time. To simplify the discussion, it is assumed that the
decoupling of matter and radiation takes place instantaneously at the
surface of last scattering. The matter of the background cosmological
model is assumed to be described by dust, \ie $p=0$. 

By the following simple argument we can approximate the interval of
dimensionless time, $\Delta\tau$, that has elapsed from the event of
last 
scattering until now. If the radiation is thermally distributed, its
energy density $\mu_{\rm r}$, as derived from the quantum statistical
mechanics of massless particles, satisfies $\mu_{\rm r} \propto T^4$
where $T$ is the temperature of the radiation (see Wald
\cite{book:Wald1984}, page 108). A non-tilted radiation fluid satisfies 
$\mu_{\rm r} \propto {\rm exp}(-4\tau)$, which implies
\begin{equation}
\frac{T_{\rm o}}{T_{\rm e}} \approx e^{-\Delta\tau}\ .
\end{equation}
Here $T_{\rm o}$ and $T_{\rm e}$ are the temperature at the present
time and at the surface of last scattering respectively. Assuming that
the process of last scattering took place when $T_{\rm e} \approx 
3000$ K, and that the mean temperature of the CMB today is $T_{\rm o}
\approx 3 $, it follows that $\Delta\tau \approx 7$. This corresponds
to a redshift of about $z\approx 1100$.

The temperature of the CMB can now be found as follows.
Introduce a future-pointing null vector ${\bf k}$
which is tangent to a light ray at a point on the CMB sky. The current
observed temperature $T_{\rm o}$ of the CMB is given by (see, for
example, Collins \& Hawking \cite{art:CollinsHawking1973a}, page 313 )
\begin{equation}
  \frac{T_{\rm o}}{T_{\rm e}} = \frac{\left(u_ak^a\right)_{\rm
      o}}{\left(u_ak^a\right)_{\rm e}} = \frac{{\cal E}(t_{\rm
      o})}{{\cal E}(\tau_{\rm e})} \ .
\end{equation}
From Eqs.\ (\ref{eq:decs}) and (\ref{eq:sgeneral}) it follows that 
\begin{equation}
  \label{eq:Texpr}
  T_{\rm o} = T_{\rm
  e}\exp\left[-\int_{\tau_{\rm 
  e}}^{\tau_{\rm o}}\left(
  1+\Sigma_{\alpha\beta}(\tau)K^\alpha(\tau)K^\beta(\tau)\right)d\tau\right]\
  . 
\end{equation}
This formula gives the temperature at time $\tau=\tau_0$ in the
direction specified by the direction cosines $K_\alpha(\tau_0)$. We
introduce angles $\theta, \varphi$ by
\begin{align}
  K_1(\tau_0) &= \sin\theta\cos\varphi\ , \nonumber \\
  K_2(\tau_0) &= \sin\theta\sin\varphi\ , \label{eq:Kobs}\\
  K_3(\tau_0) &= \cos\theta \nonumber \ ,
\end{align}
to describe positions on the celestial sphere. Note that to obtain a
correspondence with the spherical 
angles defining the direction in which an observer measures the
temperature of the CMB, one has to make the transformation $\theta
\rightarrow \pi - \theta$, $\varphi\rightarrow\varphi+\pi$. In this
way, 
$T_{\rm o}$ is a function of the angles $\theta$ and $\varphi$, \ie
\begin{equation}
  T_{\rm o} = T(\theta, \varphi)\ ,
\end{equation}
which we call the {\em temperature function} of the CMB.

The anisotropy in the CMB temperature can be described using multipole
moments (see for example Bajtlik \etal
\cite{art:Bajtliketal1985}). The  
fluctuation of the CMB temperature over the celestial sphere is
written as a spherical harmonic expansion, 
\begin{equation}
  \frac{\Delta T}{T}\left(\theta,\varphi\right) =
  \frac{T(\theta,\varphi)-T_{\rm 
  av}}{T_{\rm av}} = 
  \sum_{l=1}^{\infty}\sum_{m=-l}^l a_{lm}Y_{lm}\left(\theta,
  \varphi\right)\ , 
\end{equation}
where $T_{\rm av}$ is the mean temperature of the CMB sky. The
coefficients $a_{lm}$ are defined by
\begin{equation}
  \label{eq:almdef}
  a_{lm} = \iint\limits_{S^2} \frac{\Delta T}{T}\left(\theta,
  \varphi\right)Y^*_{lm}\left(\theta, \varphi\right)d\Omega\ ,
\end{equation}
where * denotes complex conjugation, and the integral is taken over
the 2-sphere (see for example Zwillinger \cite{book:Zwillinger1996},
pages 492-493).
The multipole moments, describing the anisotropies in a coordinate
independent way, 
are defined as 
\begin{equation}
  \label{eq:alsum}
  a_l = \left(\sum_{m=-l}^l \left| a_{lm} \right|^2\right)^{1/2}\ .
\end{equation}
The dipole, $a_1$, is interpreted as describing the motion of the
solar system with respect to the rest frame of the CMB. Therefore, the
lowest multipole moment that 
describes true anisotropies of the CMB temperature is the
quadrupole moment, $a_2$. Current observations provide an estimate for
$a_2$ as well as for the octupole moment $a_3$ (see
Stoeger \etal \cite{art:Stoegeretal1997}).

In order to compute $T(\theta,\varphi)$ and the multipole moments
$a_2$ and $a_3$ for a particular cosmological model, one has to
specify the dimensionless state, ${\bf x}(\tau_{\rm o}) = {\bf 
  x}_{\rm o}$, of the model at the time of observation, $\tau_{\rm
  o}$, and the direction of reception $K_\alpha(\tau_{\rm o})$, which
determines the angles $\theta$ and $\varphi$ on the celestial sphere
via (\ref{eq:Kobs}). The solution ${\bf x}(\tau)$, $K_\alpha(\tau)$ of
the extended system of equations (\ref{eq:feq}) and (\ref{eq:geeq}),
determined by the initial conditions ${\bf x}(\tau_{\rm o})$ and
$K_\alpha(\tau_{\rm o})$, is substituted in (\ref{eq:Texpr}), which
determines the temperature function $T\left(\theta,\varphi\right)$. The
multipoles $a_2$ and $a_3$ are then calculated by integrating over the
2-sphere (see (\ref{eq:almdef}) and (\ref{eq:alsum})). In this way
the multipole moments can be viewed as functions defined 
on the dimensionless gravitational state space, with the time elapsed 
$\Delta\tau$ since last scattering as an additional parameter: 
\begin{equation}
  a_2 = a_2({\bf x}_{\rm o}; \Delta\tau)\ , \quad a_3 = a_3({\bf
  x}_{\rm o}; 
  \Delta\tau)\ .
\end{equation}
The extended equations can be used in three ways to obtain information
about $T(\theta, \varphi)$ and the multipoles $a_2$ and $a_3$, as
follows.
\begin{enumerate}
\item Apply dynamical systems methods to the extended equations to
  obtain qualitative information about the null geodesics and the
  shear, and hence about the temperature pattern of the CMB.
\item Linearize the extended equations about an FL model, and if
  possible solve them to obtain approximate analytical
  expressions for $T(\theta, \varphi)$, $a_2$ and $a_3$, which are then
  valid for $\Sigma << 1$.
\item Use the full non-linear extended equations to do numerical
  simulations, calculating $T(\theta, \varphi)$, $a_2$ and $a_3$ for a
  given point ${\bf x}_{\rm o}$ in the gravitational state space, not
  necessarily satisfying $\Sigma_{\rm o} << 1$. As described above,
  one calculates the value of $T\left(\theta, \varphi\right)$ at each
  point of a grid covering the celestial sphere and then integrates
  numerically over the sphere to obtain $a_2$ and $a_3$. 
\end{enumerate}
One can use the observational bounds on $a_2$ and $a_3$, together with
the 
results of ii) and iii) above, to determine bounds 
on the anisotropy parameters associated with the shear
and the Weyl curvature, $\Sigma_{\rm o}$ and ${\cal W}_{\rm o}$. A
necessary condition for the model to be 
close to FL at the time of observation is that $\Sigma_{\rm o}$ and
${\cal W}_{\rm o}$ are small (see Nilsson \etal
\cite{art:Nilssonetal1999apj}).

We are in the process of applying the above methods to SH models of
various Bianchi group types. Preliminary results on Bianchi VII$_0$
models are given in Nilsson \etal \cite{art:Nilssonetal1999apj}, where it
is shown that the observational bounds on $a_2$ and $a_3$ do not
necessarily imply that the Weyl parameter ${\cal W}_{\rm o}$ is
small. Thus, an almost isotropic CMB temperature {\em does not} imply
an almost isotropic universe. An advantage of the above
methods is that they are not restricted to those Bianchi types that are
admitted by the FL models. For example, we are studying the diagonal
class B models of Bianchi type VI$_h$ as described by 
(\ref{eq:typeVIhequations}). The 
current state ${\bf x}_{\rm o}$ can be described by $\Sigma_{\rm o}$,
$k$ and $\Omega_{\rm o}$, and so the expression for the quadrupole has
the general form
\begin{equation}
  a_2 = a_2(\Sigma_{{\rm o}},k,\Omega_{\rm o}; \Delta\tau)\ .
\end{equation}
When one uses the method ii) and linearizes about the open FL model
($\Sigma_{\rm o}=k=0$), one obtains a formula of the form
\begin{equation}
  a_2 \approx \Sigma_{\rm o}I_2(\Omega_{\rm o};\Delta\tau) +
  kJ_2(\Omega_{\rm o};\Delta\tau)\ ,
\end{equation}
where $I_2$ and $J_2$ are expressed as integrals. This formula, which
is valid  
for $\Sigma_{\rm o}<<1$ and $k<<1$, leads to bounds on the shear
parameter $\Sigma_{\rm o}$ that are much weaker than those obtained in
Bianchi types I and V (see Lim \etal \cite{art:Limetal1999}). This
result shows 
that the bounds obtained for the anisotropy parameters in
Bianchi type I and V models (see, for example Bajtlik \etal
\cite{art:Bajtliketal1985}), which seem to have been taken for granted 
as being typical, are misleading. 

\section{Discussion}
\label{sec:disc}

In this paper we have shown that in the case of spatially
homogeneous models, the field equations can be augmented with the
geodesic equations, producing an extended set of first-order evolution
equations whose 
solutions describe not only the evolving geometry but also the
structure of the geodesics. Examples of the dynamics of geodesics
in some 
self-similar models, and in a simple non-self-similar model, in order
to show the predictive power of the approach, were given. We 
also made some conjectures about the qualitative behavior of geodesics
towards 
the initial Mixmaster singularity in Bianchi IX models by considering
the 
geodesics structure of the Kasner models. In light of the chaotic
nature of the Mixmaster singularity, it is expected that the geodesics
will also have some sort of chaotic behavior. To confirm
these 
speculations, we point out that further analytic studies and 
thorough numerical investigations are 
needed. We believe that the formulation of the combined field
equations and geodesic equations of this paper is very well suited for
this purpose. 

The most physically interesting aspect of the extended set of
equations is the possibility of shedding light on one of the
fundamental questions concerning the CMB, namely, does a highly
isotropic CMB temperature imply that the universe can accurately be
described by an FL model? As mentioned in section \ref{sec:temp},
the preliminary indications are that the situation is less clear-cut
than was previously thought.

We now conclude by listing some related research topics. One can easily
generalize the present formalism to spatially 
self-similar models and it is also easy to include other sources,
\eg, two non-interacting fluids, a cosmological constant, or
magnetic fields. It should 
also be possible to extend the system to include polarization. It
would also be interesting if the current formulation could be
generalized to facilitate a study of the temperature of the CMB
for density perturbed models or models with closed topologies.

\appendix
\section{Obtaining the geodesics}
For many physical purposes in SH cosmology, it is sufficient to know the
tangent vector field $\bf k$ of the geodesics. Nevertheless, it might
be interesting to find the geodesics themselves. To do so, 
spacetime coordinates $x^\mu$ ($\mu = 0,1,2,3$) must be introduced.
Once this is done, a geodesic can locally be described as a curve 
$x^\mu(\lambda)$, where $\lambda$ is the affine parameter of the geodesic.
With an SH geometry, it is natural to adapt the coordinates to the 
structure imposed by the SH condition. Expressed in terms of coordinates, 
the orthonormal frame can then be written as
\begin{equation}\label{eq:31tet}
 {\bf e}_{0} = 
 N(t)^{-1}\,(\mbox{\boldmath$\partial$}_{t}
                 -{N}{}^i(t,{\bf x})\,{\bf E}_{i}({\bf x}))
  \ , \quad
 {\bf e}_{\alpha} =  {e}_\alpha{}^{i}(t)\,{\bf E}_{i}({\bf x})\ ,
\end{equation}
where $N$ and $\vec{\bf N} = N{}^{i} {\bf E}_i$ are the lapse function and
the shift vector field respectively (for restrictions on the shift
vector field, see Jantzen \& Uggla \cite{art:JantzenUggla1999}). 
The spatial frame
$\{ {\bf  E}_i({\bf x}) \}$, where $i = 1,2,3$ (in this appendix, and
only here, latin indices $i,j,k...=1,2,3$ denote spatially homogeneous
time independent frame indices since it is only here that this type of
frame is used), tangent to
each hypersurface, is not 
only invariant under the action of the Bianchi symmetry group
but has structure or commutator functions ${C}{}^k\!_{ij}$ 
which are constants throughout the spacetime,
defined by 
\begin{equation}
 \left[\, {\bf E}_i, {\bf E}_{j}\,\right]
 = {C}^{k}\!_{ij}\,{\bf E}_{k} \ .
\end{equation}

It is
possible to construct the orthonormal frame explicitly in terms of
local coordinates $\{t,x^i\}$ ($i=1,2,3$) adapted to the SH
hypersurfaces. The spatial frame 
$\{ {\bf E}_i \}$ is characterized by the Lie dragging condition 
 ${\cal L}_{{\boldsymbol{\partial}_t}} {\bf E}_i =0$ 
which implies the time independent local coordinate expression 
${\bf E}_{i} 
= {E}_{i}{}^{j}(x^{k})\,{\boldsymbol{\partial}}_{j}$ 
for the invariant spatial frame.
Explicit coordinate expressions for 
${E}_{i}{}^{j}(x^{k})$
follow from the representation of the left invariant vector fields in 
canonical coordinates of the second kind in Jantzen
\cite{art:Jantzen1979,art:Jantzen1984}. The  
relation between the orthonormal frame components of the tangent
vector  
field to the geodesics and its corresponding coordinate components
yields the relations
\begin{mathletters}
  \label{eq:l1l2}
  \begin{eqnarray}
    \frac{d\lambda }{d\tau} &=& (H {\cal E})^{-1}\ , \label{eq:l1}\\
    \frac{dx^i}{d\tau} &=& \sum_j[ -N^j(\tau,{\bf x}) + 
    H^{-1}(\sum_\alpha K^\alpha(\tau) e_\alpha{}^j(\tau))] E_j{}^i({\bf
      x})\ ,  
    \label{eq:l2} 
  \end{eqnarray}
\end{mathletters}
where the $K^\alpha$ are the energy-normalized components of the
geodesic tangent vector field.
Here one can choose an automorphism adapted shift in order to set certain 
components of $e_\alpha{}^j(\tau)$ to zero, see
Jantzen \& Uggla \cite{art:JantzenUggla1999}.
One usually sets the shift to zero. In this latter case one has to add
equations governing some of the $e_\alpha{}^j(\tau)$ components. These 
are obtained from the commutator relations and are consequences of 
the zero-shift gauge. We will not do this for the general case. Instead 
we will look at an example.

\subsection{An example: Non-tilted class A models with zero shift}
In this case, with zero shift, equation (\ref{eq:l1l2})
becomes
\begin{align}
  d\lambda/d\tau &= (H {\cal E})^{-1}\ , \\
  dx^i/d\tau &= \sum_\alpha H^{-1} K^\alpha e^{-{\beta_\alpha}} 
  E_\alpha{}^i(x)\ ,
\end{align}
where $e_\alpha\!^j = \delta_\alpha^j{\rm e}^{-\beta_\alpha}$.
The commutator equations (see Jantzen \& Uggla
\cite{art:JantzenUggla1999} or WE, 
chapter 10) yields 
\begin{align}
  d\beta_{1,2}/d\tau &= 1 + \Sigma_+ \pm \sqrt{3}\Sigma_-\ , \\
  d\beta_3/d\tau &= 1 - 2\Sigma_+\ . 
\end{align}
The next step is to introduce $H^{-1}{\rm e}^{-\beta_\alpha}$ as new
variables, but note that they are not independent of 
$N_\alpha$ (see WE, chapter 10). This
results in the complete extended system of equations, which is
necessary in order to obtain the individual geodesics.

\appendixonfalse

\section*{Acknowledgements}
We thank Woei Chet Lim for helpful discussions and for commenting in
detail on an earlier draft of the manuscript. This research was
supported in part by a grant from the 
Natural Sciences \& Engineering Research Council of Canada (JW),
the Swedish Natural Research Council (CU), G{\aa}l\"ostiftelsen (USN),
Svenska Institutet (USN), Stiftelsen Blanceflor (USN) and the
University of Waterloo (USN).


\begin{thebibliography}{10}

\bibitem{book:WainwrightEllis1997}
Wainwright, J. and Ellis, G. F.~R.  (1997). {\em Dynamical systems in
  cosmology}  ({C}ambridge {U}niversity {P}ress, Cambridge).

\bibitem{art:EllisMacCallum1969}
Ellis, G. F.~R. and MacCallum, M. A.~H.  (1969). {\em Commun. Math. Phys.} 12,
  108.

\bibitem{art:CollinsHawking1973a}
Collins, C.~B. and Hawking, S.~W.  (1973). {\em Mon. Not. Roy. Astr. Soc.} 162,
  307.

\bibitem{art:Barrowetal1983}
Barrow, J.~D., Juszkiewicz, R., and Sonoda, D.~H.  (1983). {\em Nature} 309,
  397.

\bibitem{art:Bajtliketal1985}
Bajtlik, S., Juszkiewicz, R., Proszynski, M., and Amsterdamski, P.  (1985).
  {\em Astrophys. J.} 300, 463.

\bibitem{art:Doroshkevichetal1975}
Doroshkevich, A.~G., Lukash, V.~N., and Novikov, I.~D.  (1975). {\em Sov.
  Astronomy} 18, 554.

\bibitem{art:Wainwrightetal1999}
Wainwright, J., Hancock, M.~J., and Uggla, C.  (1999). {\em Class. Quant.
  Grav.} 16, 2577.

\bibitem{art:CollinsStewart1971}
Collins, C.~B. and Stewart, J.~M.  (1971). {\em Mon. Not. Roy. Astr. Soc.} 153,
  419.

\bibitem{art:BKL}
Belinskii, V.~A., Khalatnikov, I.~M., and Lifschitz, E.~M.  (1970). {\em Adv.
  Phys.} 19, 525.

\bibitem{art:ColeyWainwright1992}
Coley, A.~A. and Wainwright, J.  (1992). {\em Class. Quant. Grav.} 9, 651.

\bibitem{book:Wald1984}
Wald, R.~M.  (1984). {\em General relativity}  ({U}niversity of {C}hicago
  {P}ress, Chicago).

\bibitem{book:Zwillinger1996}
Zwillinger, D.  (1996). {\em CRC Standard mathematical tables and formulae}
  (CRC Press, Boca Raton).

\bibitem{art:Stoegeretal1997}
Stoeger, W.~R., Araujo, M.~E., and Gebbie, T.  (1997). {\em Astrophys. J.} 476,
  435.

\bibitem{art:Nilssonetal1999apj}
Nilsson, U.~S., Uggla, C., and Wainwright, J.  (1999). {\em Astrophys. J.
  Lett.} 522, L1.

\bibitem{art:Limetal1999}
Lim, W.~C., Nilsson, U., and Wainwright, J.  (1999). ``The temperature
of the cosmic microwave background in Bianchi VI$_h$ universes'', in
preparation. 

\bibitem{art:JantzenUggla1999}
Jantzen, R.~T. and Uggla, C.  (1999). {\em J. Math. Phys.} 40, 353.

\bibitem{art:Jantzen1979}
Jantzen, R.~T.  (1979). {\em Commun. Math. Phys.} 64, 211.

\bibitem{art:Jantzen1984}
Jantzen, R.~T. (1984). ``Spatially homogeneous dynamics: a unified
picture'', In 
  Ruffini, R. and Melchiorri, F. editors, {\em Proc. {I}nt. {S}ch. {P}hys. 'E
  Fermi {C}ourse {LXXXVI} on 'Gamov cosmology}, page~61,
(Amsterdam:North Holland). 

\end{thebibliography}

\end{document}